\documentclass[twocolumn]{article} 

\usepackage{arxiv}

\usepackage[all]{nowidow}

\usepackage[utf8]{inputenc} 
\usepackage[T1]{fontenc}  
\usepackage{hyperref}       
\usepackage{url}            
\usepackage{booktabs}       
\usepackage{amsfonts}       
\usepackage{microtype} 

\usepackage{float}  
\usepackage{graphicx}
\usepackage[labelfont=bf]{caption}
\usepackage{tcolorbox} 
         
\usepackage{enumitem}                  
\usepackage{enumerate}

\newcommand{\gap}{personal informatics analysis gap~}
\newcommand{\gapNS}{personal informatics analysis gap}

\newcommand{\etal}{et al.~}

\renewenvironment{quote}%
  {\list{}{\leftmargin=0.1in\rightmargin=0in}\item[]}%
  {\endlist}

\title{Exploring the personal informatics analysis gap: \\ ``There's a lot of bacon''}

\def \shortauthor {Moore et al.}
\author{
  Jimmy Moore \\
  University of Utah\\
    \texttt{jimmy@cs.utah.edu}
    \And
  Pascal Goffin \\
  Asvito Digital AG \\
     \texttt{ppjgoffin@gmail.com}\\
    \And
  Jason Wiese \\
  University of Utah \\
    \texttt{wiese@cs.utah.edu}
    \And
  Miriah Meyer \\
  University of Utah\\
  \texttt{miriah@cs.utah.edu}
}

\begin{document}
\twocolumn[
  \begin{@twocolumnfalse}
    \maketitle

\begin{abstract}
Personal informatics research helps people track personal data for the purposes of self-reflection and gaining self-knowledge. This field, however, has predominantly focused on the data collection and insight-generation elements of self-tracking, with less attention paid to flexible data analysis. As a result, this inattention has led to inflexible analytic pipelines that do not reflect or support the diverse ways people want to engage with their data. This paper contributes a review of personal informatics and visualization research literature to expose a gap in our knowledge for designing flexible tools that assist people engaging with and analyzing personal data in personal contexts, what we call the \textit{personal informatics analysis gap}.  We explore this gap through a multistage longitudinal study on how asthmatics engage with personal air quality data, and we report how participants: were motivated by broad and diverse goals;  exhibited patterns in the way they explored their data;  engaged with their data in playful ways; discovered new insights through serendipitous exploration; and were reluctant to use analysis tools on their own. These results present new opportunities for visual analysis research and suggest the need for fundamental shifts in how and what we design when supporting personal data analysis.
\end{abstract}

\vspace{2mm}


\vspace{8mm}

  \end{@twocolumnfalse}
]

\begin{tcolorbox}[floatplacement=!b,float,left=2mm,right=2mm,top=1mm,bottom=1mm]
\small
This is the authors' preprint version of this paper. License: CC-By Attribution 4.0 International. Please cite the following reference: \\
Jimmy Moore, Pascal Goffin, Jason Wiese, Miriah Meyer.
Exploring the personal informatics analysis gap: ``There's a lot of bacon''. 
\textit{TVCG Special Issue on the 2021 IEEE Visualization Conference (VIS)}, to appear, 2021.
\end{tcolorbox}

\section{Introduction}\label{sec:introduction}

Personal informatics research focuses on tools and technologies that help people to collect and make sense of personally relevant data for the purposes of self-reflection and gaining self-knowledge \cite{li2010stage}. The field builds on increasingly available and ubiquitous technology for capturing data about our everyday lived experience, and is driven by the idea that data will help us make better, smarter decisions about how we live our lives.
Personal informatics researchers study how and why people engage with personal data in a broad range of application areas, including fitness~\cite{consolvo2006design,consolvo2008activity,epstein2014taming,mauriello2014fitness}, indoor air quality \cite{Moore2018,kim2009inair,kim2010inair,kim2013inair,fang2016airsense,Gupta2010,Campbell:2010:WMS:1864349.1864378}, health \cite{tsai2007usability,chung2019identifying,chung2017personal}, and time management \cite{peesapati2010pensieve,lindqvist2011m}.

Researchers have characterized the stages people go through when collecting and engaging with their personal data \cite{li2010stage,epstein2015lived}, in part to classify and assess research efforts across them. A recent survey of over 500 personal informatics publications shows, however, that the field focuses more on designing tools that support data collection and reflection, and less on tools to support sense-making and analysis \cite{epstein2020mapping}. Existing personal informatics tools that \textit{do} support analysis do so in a specific way: they remove the analysis burden of data processing and task operationalization by baking-in specific analysis workflows. These workflows are designed by personal informatics researchers \cite{Huang2015}, not the people who are collecting and analyzing their personal data, resulting in tools that do not always reflect the ways people want to engage with or think about their data \cite{epstein2014taming}.  As a result, a gap exists between the capabilities personal informatics tools support for helping people to engage with personal data, and the range of tasks people want to perform on the data they collect. 

This gap presents an opportunity for the visualization community, who have already thought deeply about ways to support people engaging with data, to make an impact in this space. Research in visual analytics has focused on supporting people working with data. This work, however, is tailored for domain experts working in professional contexts \cite{wong2018towards,sedlmair2012design} who bring motivations, skills, and experiences to data analysis that cannot be assumed about people in personal contexts \cite{Huang2015}. On the other hand, personal visual analytics has emerged to unify several threads of visualization research that ``empower everyday people through exploring data'' within personal contexts \cite{Huang2015}. Despite consolidating several research areas focused on nonprofessional contexts, most approaches are not designed to support in-depth analysis of personal data.  This subfield is instead primarily oriented toward promoting data awareness, exploration, or social sharing.

We refer to the lack of attention to flexible analytic tools for supporting people to engage with their personal data as the \textit{\gapNS}. We explore this gap using a new interview method \cite{Moore2021} that we conducted as part of a multistage longitudinal study on how asthmatics engage with personal air quality data \cite{Moore2018}.
This work reports on what we found when probing the \gap within the context of our longitudinal study: participants were motivated by broad and diverse goals;  exhibited patterns in the way they explored their data;  engaged with their data in playful ways; discovered new insights through serendipitous exploration; and were reluctant to use analysis tools on their own. These results expose new opportunities for visualization research and suggest the need for fundamental shifts in how and what we design for when supporting personal data analysis.

The specific contributions of this work are twofold. First, we identify the existence of the \gap through a review of personal informatics and visualization literature, and validate it through our experiences of attempting to design visual analysis tools in our longitudinal study. Second, we offer three design considerations for the visualization community to help bridge this gap, informed by the results of data engagement interviews we conducted with our participants. These contributions point to new opportunities for the visualization community to design for people engaging with their personal data.

In the next section, we characterize the \gap in more detail, drawing on our review of the personal informatics and visualization literature.  We describe our own journey encountering and exploring the \gap in the context of a longitudinal study with seven asthmatic families in \autoref{sec:methods} and outline our findings from observing how participants engaged with their data in \autoref{sec:results}.  We propose three design recommendations for exploring this space in \autoref{sec:discussion}, and  conclude with ideas for future work in \autoref{sec:conclusion}.

\section{Identifying the \gap}\label{sec:related_work}

In this section, we review literature from the personal informatics and visualization communities, with a focus on how each contributes to our understanding of supporting people engaging {with} their personal data. Through this literature analysis,  we observe gaps in knowledge and argue that the culmination of these constitutes the \gapNS.

\subsection{Personal informatics tools}

Personal informatics is a research area in human-computer interaction that focuses on the ``tools and technologies that help people to collect personally relevant data for the purposes of self-reflection and gaining self-knowledge'' \cite{li2010stage}. Interpreting and reflecting on personal data in any meaningful way requires a deep, contextual awareness of people's lives. This knowledge is critical for drawing interesting conclusions or insightful discoveries. Those lacking the same situated knowledge of someone's social contexts, routines, and priorities are unable to correctly interpret their logged personal data, much less meaningfully analyze or make recommendations from them \cite{tolmie2016has,fischer2016just}.    
    
Personal informatics emerged from the quantified self movement with early tools designed to help users gain self-insights by tracking single facets of their lives, such as their diet \cite{siek2006we} or physical activity \cite{lin2006fish}. Subsequent research found that reflecting on single data streams limited the kinds of insights people could derive from their data \cite{Bentley2013}.  Thus, systems for tracking multiple facets of people's lives emerged to help sustain user engagement and improve insight generation \cite{li2010stage,haddadi2014quantified,epstein2014taming, wiese2017}. This shift, however, imposes a greater analytic complexity that users find unmanageable \cite{choe2014understanding, jones2015exploring, jones2018dealing}.  
Attempts to curb this complexity either seek to prioritize smaller, more manageable tracking tasks as a way to narrow the design and problem spaces, or implement analytic pipelines that automatically detect and present potentially interesting correlations based on statistical analysis~\cite{jones2015exploring, jones2018dealing}.  These solutions involve fixed processing pipelines that give the user little control over the kinds of analysis or types of insights afforded to them ~\cite{Huang2015}, and result in frustration with receiving obvious insights \cite{Bentley2013}, or having to review too many potential correlations \cite{jones2018dealing}.  
    
One step toward supporting more interactive and exploratory personal data analysis involves a technique known as data \textit{cuts} \cite{epstein2014taming}.  Cuts refer to subsets of data, chosen by underlying features, that support detailed and potentially interesting comparisons. In building a tool to support analysis of cuts, Epstein \etal developed a set of predefined cuts based on a survey of the kinds of questions users had of their personal data. User evaluations of these cuts found them to be effective at supporting \textit{some} of their questions, but the predefined cuts failed to meet everyone's needs or preferences for engaging their data. These findings echo those in other studies that attempt to automate insights \cite{jones2018dealing,Bentley2013}, and led the authors to recommend that  "designs do not attempt to limit cuts based on stated goals and instead offer a variety of cuts" \cite{epstein2014taming}. What remains unclear, however, is how to design for arbitrary and flexible analysis, especially in the context of open-ended or ill-defined goals.
    
\begin{figure}
        \centering
        \includegraphics[width = \columnwidth]{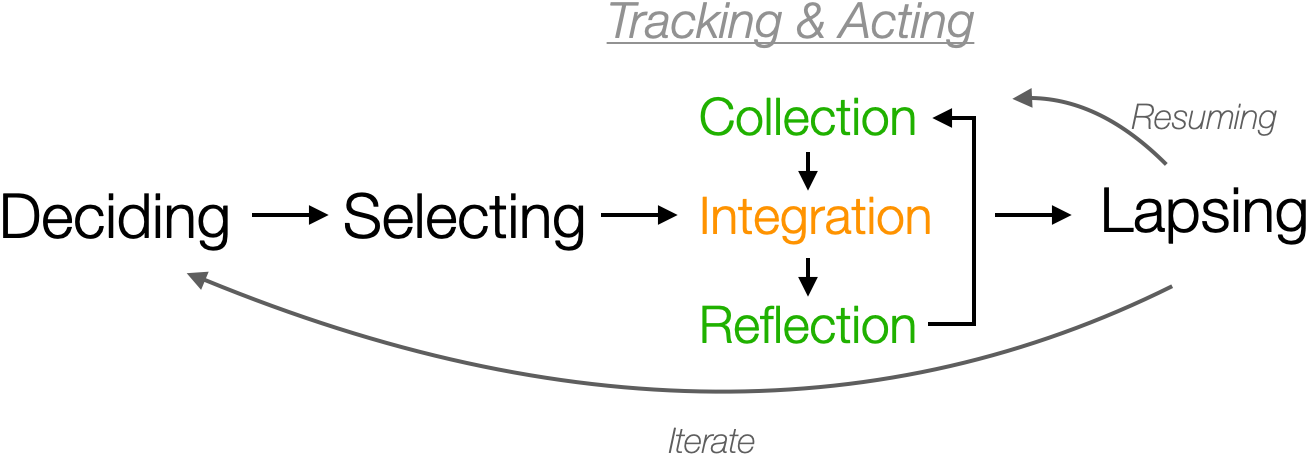}
        \caption{ The lived informatics model  of personal informatics \cite{epstein2015lived}.  A recent survey of over 500 personal informatics publications \cite{epstein2020mapping} found researchers focus primarily on the collection and reflection stages (green) of the tracking and acting cycle , and less on the integration stage (orange), where data is combined, transformed, and analyzed.}
        \label{fig:lived_informatics_model}
    \end{figure}

Alongside system design, personal informatics research has also developed models to describe the ways and reasons people track personal data \cite{li2010stage,epstein2015lived}. The stage-based model for personal informatics systems \cite{li2010stage} classifies how people self-track in practice, and outlines an iterative five-stage process for users pursuing goal-oriented behavior change: \textit{preparation}, \textit{collection}, \textit{integration}, \textit{reflection}, and \textit{action} stages.  Later, the lived informatics model \cite{epstein2015lived} revised the stage-based model to reflect a more inclusive classification of people's tracking behaviors and motivations (\autoref{fig:lived_informatics_model}). Specifically, it acknowledges users' decisions of whether and how to track personal data, their ability to lapse and resume self-tracking practices, and to track or review personal data for non-goal-oriented motivations. These models are foundational in the personal informatics community, and guide how researchers develop and study systems that help people improve aspects of their lives.
    
Personal informatics research, however, is not uniformly distributed over these model stages.  A recent retrospective survey of over 500 personal informatics publications found considerable research that addresses barriers in the collection and reflection stages, but with significantly less focus on integration \cite{epstein2020mapping}  (\autoref{fig:lived_informatics_model}). 
The integration stage remaining understudied --- the stage where data are combined, transformed, and analyzed ---  echoes other findings on the inherent difficulties with supporting these functionalities \cite{jones2018dealing}. Tackling this challenge requires more attention and research efforts to raise greater awareness on these disparities; this work is a step in that direction.
   
\vspace{0.25cm}
\noindent \textbf{OBSERVATION 1:} Personal informatics leverages people’s deep contextual knowledge to collect and reflect on personal data, but requires further study to develop flexible analysis tools that empower people to engage with their personal data in unique ways.

\subsection{Flexible visual analysis tools}
    
Visualization research has a long history of developing tools to help people productively engage with data, but largely targets those working in professional contexts. Developing visual analysis tools for use in professional settings stems from two core threads of research: the design of bespoke tools for domain experts, and the creation of powerful and flexible systems for data analysts.
    
A proliferation of bespoke tools for supporting rich visual analysis across a broad range of fields stems from the visualization research community's embrace of a call for ``collaborating closely with domain experts who have appropriate driving tasks in data-rich fields to produce tools and techniques that solve clear real-world needs''~\cite{munzner2006nih}. 
The dominant research approach behind the design and development of these tools is \textit{visualization design study}, an approach to problem-driven visualization research that emphasizes designing visual analysis tools in close collaboration with domain experts~\cite{sedlmair2012design}. 
Design study is now a standard method for conducting visualization research inquiry, informed by validation methods \cite{munzner09nested_model}, process models \cite{sedlmair2012design,Mckenna2014, mccurdy2016action}, rigor criteria \cite{meyer2019criteria}, and guiding scenarios \cite{sedlmair16}. 
Published research papers reporting on design studies cover a range of application areas, but they all focus on developing analysis tools for domain experts \textit{working} with data. None report on collaborations outside a professional context. 
    
Research into visualization systems for data analysts arises from the significant increase in the number of professionals whose primary task is data analysis. Various interview studies focus on how data analysts do their work, and offer characterizations of their overall analysis process within the organizational context~\cite{kandel2012enterprise}; their patterns of exploratory data analysis~\cite{Alspaugh2019}; their impediments to efficient data analysis~\cite{kandogan2014data}; their roles within software development teams~\cite{kim2016emerging}; and their unique considerations when working with cloud architectures~\cite{fisher2012interactions}. These studies extend and modernize earlier research on intelligence analysts' work practices~\cite{pirolli2005sensemaking,cowley2005glass,patterson2001predicting,wright2006sandbox} by seeking to understand the sense-making and information-foraging processes of data analysts ~\cite{pirolli1999information,russell1993cost}.
   
The results of these studies inform a growing ecosystem of tools for data wrangling~\cite{kandel2011wrangler,bigelow2019origraph},  interactive visual analysis~\cite{bostock2011d3,mackinlay1986automating}, and visualization recommendations~\cite{wongsuphasawat2015voyager,wongsuphasawat2016towards,mackinlay2007show}. 
    
These tools for professionals require users to translate their questions into accompanying analysis tasks and make appropriate decisions based on visualizations that they see~\cite{amar2005knowledge,chul2011visual}. People without these skills, however, struggle to use visualization tools effectively~\cite{law2018duet,grammel2010information,chul2011visual}. Generally speaking, people engaging with data in personal contexts tend to have less time, training, patience, motivation, capabilities, and crisply actionable tasks for their personal data than do professional analysts \cite{Huang2015}. 
    
\vspace{0.25cm}
\noindent \textbf{OBSERVATION 2: } Visual analytics research excels at designing flexible data analysis tools for experts working in professional contexts, but does not target personal contexts, where existing tools do not easily transfer to the needs, skills, and motivations of people exploring their own data.

\subsection{Everyday visual analysis}
    
Visualization's growing use and consumption in everyday contexts has spurred new research into making data accessible and understandable to everyone. Everyday visualization encompasses multiple use-cases and research goals, with early work in this space exploring ways to engage people with new visualization techniques \cite{wattenberg1999visualizing, wattenberg2005baby}, democratize visualization \cite{viegas2007manyeyes,heer2009voyagers,danis2008your,heer2005vizster}, and empower people through exploring data \cite{huron2014constructive, Huang2015, danziger2008information}.  This community has since expanded its scope to investigating physicalizations for awareness or goal-setting \cite{Khot2015,Khot2013,Stusak2014, thudt2018self,Thudt2016}, engagement with personal data in the home \cite{Houben2016}, and even emotional connection to personal data when mapped to living artifacts \cite{holstius2004infotropism,Botros2016}.
    
Engaging with data in personal contexts fundamentally differs from engagement in more professional contexts encountered in standard visual analytics research.   The subfields of personal visualization and personal visual analytics emerged to formalize the  \textit{personal context}, identifying the  ``different motivations, priorities, role expectations, environments, or time and resource budgets as compared to professional situations'' \cite{Huang2015}.   This distinction was made to unite largely independent research communities within visualization and personal informatics that, by virtue of focusing on nonprofessional situations, extends to cover a broad range of use-cases and data scopes. Huang \etal \cite{Huang2015} recognize that this breadth ``subsumes many related fields'', yet, despite its size, recent work identifies a lack of significant activity in this space.  In a recent article on reaching broader audiences with data visualization, Lee \etal  acknowledge that ``only a limited number of researchers have continued to work at the intersection of visualization and personal informatics''\cite{Lee2020}. 
    
Although recent work incorporates personal agency into visual interaction mechanisms \cite{Koytek2018}, or explores proof-of-concept tools to support self-reflection in controlled lab studies \cite{Aseniero2020}, it does not incorporate participants' personal data as a part of the analysis process.  The overwhelming majority of research occurring within personal contexts remains rooted in exploration, awareness, or social sharing. Everyday visualization has yet to deploy truly flexible, scalable, or in-depth data analysis capabilities for personal data.
 
\vspace{0.25cm}
\noindent \textbf{OBSERVATION 3:} The democratized analysis goals of everyday visualization aim to empower people to explore data, but this field has yet to design for flexible, in-depth analysis of personal data

\subsection{The \gap}

We characterize the \gap as a \textbf{lack of focused research and design of systems that allow people to flexibly analyze their personal data}. Any solution that overcomes this gap must acknowledge and incorporate people's personal and contextual knowledge, support flexible analysis that covers a variety of different circumstances and goals, and empower people to deeply and richly engage with their personal data.
Although the fields of personal informatics, visual analytics, and everyday visualization contribute individual strengths toward bridging this gap, no one field has yet to focus on it:
    
    \begin{itemize}[nosep]
        \item Personal informatics research excels at developing systems that support users to collect and reflect on their personal data, but lacks flexible analysis tools for personal data.
        \item Visual analytics research excels at developing flexible and customized analysis tools, but these tools are tailored for professional contexts that are difficult to transfer to personal contexts.
        \item Everyday visualization excels at empowering people to explore data in personal contexts, but has yet to prioritize systems that support in-depth analysis of personal data.
    \end{itemize}

Enlisting each field's strength affords an opportunity for developing new approaches to learn how people engage with personal data, and for designing new tools and systems that will support them in doing so. Personal informatics' experience with collecting and acting on personal data can be augmented by visual analytics' background in customized analysis environments, and brought together with the everyday analysis goal of personal empowerment. 

Uniting these fields will help researchers consolidate existing design knowledge, experience, guidance, and methods to more effectively design visual analysis tools that support people engaging with their personal data. In the rest of this paper, we report on our explorations of the \gap that drew upon expertise from each of these fields during a three-year longitudinal study of asthmatic families.

\section{Methods}\label{sec:methods}

    \begin{figure*}
        \centering
        \includegraphics[width = \textwidth]{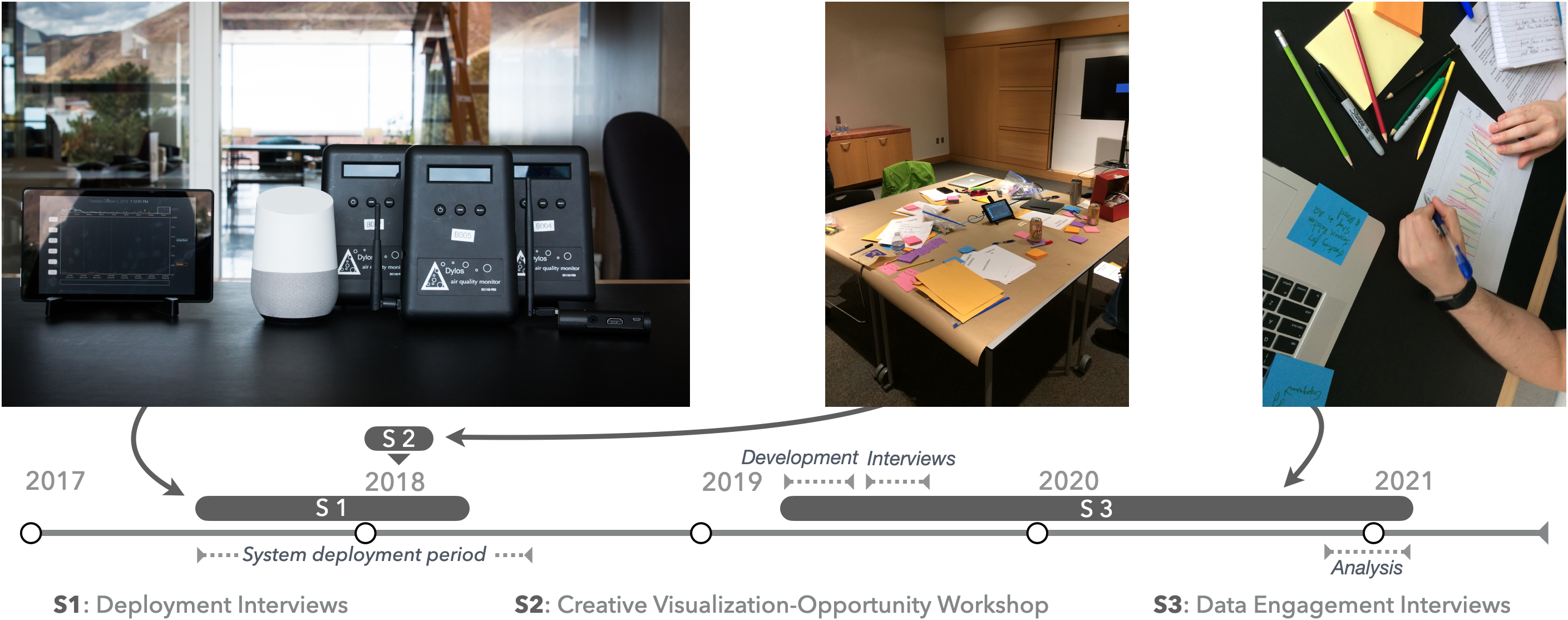}
        \caption{Longitudinal study timeline. Research stage S1 involved six long-term field deployments of a wireless air quality sensing system \cite{Moore2018} to determine how participants would interact with it over time.  Stage S2 applied a creative visualization-opportunities workshop \cite{kerzner2018framework} to collect further information on what participants wanted to do with their data.  We conducted data engagement interviews \cite{Moore2021} in stage S3 to observe how participants engaged with their data.}
        \label{fig:study_timeline}
    \end{figure*}

Our own journey exploring the \gap unfolds over a three-year longitudinal study working with asthmatic study participants (\autoref{fig:study_timeline}). This study explored how data about the air quality in the participants' homes might support them in making changes to their daily activities to improve their respiratory health. In the first research stage (S1), six families each received a wireless air quality sensing system consisting of three air quality monitors that participants could place in and around their home. These sensors measured and logged the concentration of airborne particulate matter of their immediate environments each minute, and streamed their data to an interactive interface hosted on a tablet device. A prior report provides additional details on the deployment and hosted system, along with evidence that developing flexible systems supported study participants engaging with their data in personal ways \cite{Moore2018}.

This first stage focused on how participants engaged with an air quality sensing system in their homes over a long-term deployment. We had planned to analyze transcripts from interviews conducted throughout the deployment in order to develop design requirements for an improved visualization interface based on the kinds of analysis the participants did, or wanted to do, during S1. Having tailored S1 for studying \textit{how} participants used their air quality measurement system, however, and not collecting data on \textit{what} they wanted to do with it, we lacked insight into how to design such a system.  To correct for this, we conducted a creative visualization-opportunities workshop \cite{kerzner2018framework} (S2) to characterize participants' goals and motivations for engaging their personal air quality data. The workshop outcomes, however, captured a diverse range of high- and low-level goals that only expanded the possible design space, instead of informing new possibilities.  Participants' breadth of questions further prevented us from knowing what it was they would do, or would know what to do, with their data. Despite having applied a range of visualization design methods to understand what our participants needed, we were no closer to a viable design. 

As we reflected on our overall approach, we began to realize that these standard practices for eliciting analysis goals and design requirements had been developed in collaboration with domain experts who work closely with data in professional contexts.  Consequently, these tactics were unsuccessful when used with our participants, whose deployments prevented them from deeply engaging with their personal air quality data, or even fully knowing what information was logged and stored.   Participants' lack of experience in engaging with their data put us at a disadvantage when applying techniques that presumed this level of collaborator knowledge.  Once we began to consider how participants' inexperience with their own data influenced our design challenges, a gap began to emerge between what we, as visualization designers, were trained to do, versus what was effective for engaging our target users. We have come to call this the \gap.   Realizing this gap --- both through literature and our own experiences --- changed our design thinking and motivated us to develop a new interview technique designed specifically to engage people with their personal data, which we call the \textit{data engagement interview} \cite{Moore2021}.  We conducted data engagement interviews  with each of our study participants in S3 and report on those interview outcomes in this work. 

\subsection{Data engagement interviews}
 
In order to explore the \gapNS, we developed a novel interview method called the \textit{data engagement interview }\cite{Moore2021}. This interview method incorporates a dedicated data analyst as part of the interview team for providing real-time visual data analysis of the participant's personal data during the interview. The interview is structured to take participants through different types of data engagement, providing researchers opportunities to observe how people operationalize their questions, which analysis tasks they engage with, and what they find interesting and informative in their data.

We conducted seven data engagement interviews in the S3 stage of our longitudinal study. In these interviews, we integrated daily EPA Air Quality Index classifications\footnote{https://www.airnow.gov/aqi/}, approved access to participants’ asthma health surveys collected through the parent medical study \cite{PRISMS}, and additional environmental data streams including ambient temperature and humidity as external contextualizing data. Joining these data streams provided a rich set of analysis opportunities for our participants to be able to contextualize their indoor air quality data streams. 

All participant interviews proceeded from semi-structured scripts to ensure meeting primary goals while affording us the flexibility to more deeply explore participants’ motivations and thought processes, as necessary. The data engagement interview protocol was cleared by our university’s institutional review board and is available in supplemental materials. All interview participants received a \$20 Amazon gift card.

\subsection{Participants}

We retained all seven study participants from S1 \cite{Moore2018} for the data engagement interviews due to their prior experience and engagement with reviewing personal indoor air quality data. These participants were themselves asthmatic (P1, P2, P4a, P5, P6) or primary caregivers to asthmatic children (P2, P3, P4), and they were chosen from a concurrent university-run medical study involving asthmatic families \cite{PRISMS}.  We included participant P4’s teenage daughter, P4a, as an additional study participant given her significant engagement throughout this study. Other participant family members also contributed feedback and suggestions during the data engagement interviews but were otherwise not involved.  

Participants were alike in that asthma had impacted their lives, but the extent and degree to which they were affected was entirely unique.  The individualized aspect of participants' asthma sensitives and severity influenced how they engaged with their data, and what they sought from it.   Participant P1, an asthmatic sensitive to pollen and other outdoor irritants,  used her air quality system as a personal planning tool to stay inside when outdoor particulate concentrations were high. Participants P3 and P4 are primary caregivers to asthmatic children and wanted to use their data to provide a healthy home environment for their family. Participants P5 and P6 are adult asthmatics who already understood their symptoms and were interested to explore how environmental factors affected their sensitivities. Participant P4a, our youngest study member, was less concerned with the health impacts of air quality and more curious to see how air quality affected those around her at a community level. Participant P2 is medically disabled as a result of her severe asthma and was interested in how the data could improve her quality of life.  Due to her health complications, however, she remained hesitant to interpret or act on any personal data.  Instead, she preferred to receive personalized advice from medical professionals on ways she could improve her living space. 

 \subsection{Data analysis}
 
The 7 data engagement interviews generated 8.9 hours of interview audio. The interview audio was transcribed and analyzed through several rounds of affinity mapping and group discussion between the authors in S3. After reporting on a number of findings in a companion paper \cite{Moore2021}, we pitched additional results in a pre-paper talk to our research group to collect feedback on our ideas. These additional results focused on implications for designing within the \gapNS. Afterwards, the authors performed an additional round of discussion to settle on the final set of results and implications presented in this work.

\section{Results}\label{sec:results}

In this section, we present results from our analysis of the data engagement interviews. These results illustrate the varied ways that participants approached, analyzed, and engaged with their personal data.

\subsection{Diverse goals and questions} \label{sec:diverse_goals}

All our participating families received the same deployment, collected the same types of measurements, and had the same interest in improving their families' respiratory health. Yet, despite these similarities, the specific questions our participants brought to their data engagement interviews, and the types of insights they acquired, were surprisingly diverse. 

Participants' data analysis goals covered a myriad of topics: 
\begin{quote}
    \textbf{P2:} What's going to be the best vacuum cleaner to help keep dust down?
\end{quote}   
\begin{quote}
    \textbf{P4a:} I'd like to compare my data with other people's data to see how people were affected by outdoor air quality, near me or in the valley.
\end{quote}
\begin{quote}
    \textbf{P5}: Can I anticipate days I shouldn't go outside?
\end{quote}
\begin{quote}
    \textbf{P6:} Should I move?
\end{quote}    

The analysis approaches the participants took were equally diverse. For example, P1's interest in reviewing her air quality data stemmed from wanting to understand whether periods of poor outdoor air quality impacted her indoor air quality. These discussions unfolded as she and her husband jointly explored data together, but became a broader conversation on indoor and outdoor air quality dynamics after witnessing how their personal activities impacted their living spaces. In contrast, P3 and P4 each chose to step through their air quality data spike-by-spike to see what annotations were associated with the most prominent outliers.  These explorations evolved into discussions about the proportions of spikes associated with particular activities or locations.  Reframing their air quality events by faceting on its underlying properties changed the way both participants came to see their indoor air quality.  Finally, in a third example, P4a was less curious about her air quality measurements and preferred to compare her weekly health survey responses against those of other participants.  This comparison served as a mechanism to determine whether her respiratory issues were more likely triggered by widespread environmental conditions -- which she presumed would equally affect other participants --  as opposed to personal behaviors, and affect only her. 

Even though we collected questions and goals from only seven participants, their diverse personal interests cover a wide range of analysis needs.  The questions  require integrating data over different locations (\textit{Q: How does outdoor air quality compare across the valley?}); timescales (\textit{Q:What tends to be the worst time for my indoor air quality?}); participants (\textit{Q: How does my indoor air quality compare to other participants?}); and data sources (\textit{Q: Can I correlate my air quality data to my health surveys?}). 
This diversity in analysis needs poses a significant challenge for designing a tool that is capable of supporting them all.

\subsection{Pattern of exploration} \label{sec:patterend_engagement}

Despite the varied data streams and integrations that our participants' diverse questions demanded, we \textit{did} see a generalized pattern in how they explored their data. This pattern was consistent regardless of their underlying motivations or enthusiasm, and often first appeared after we brought up an overview of their data early in the interview:
\begin{quote}
    \textbf{Interviewer}:  What would you like to start with?  What's the first thing you want to see?\\
    \textbf{P2:} I'd like to just ... okay, let's glance at this great, big, huge, orange [spike]. Did we mark what that was? If so, what was it?
\end{quote}
Large and visibly prominent spikes were a common distraction when reviewing participant data, and one of the most frequently requested features to explore:
\begin{quote}
    \textbf{P3:} I think I would start with the bad peaks. I mean like this [spike] right here, this [high magnitude spike]. Some of these higher ones.  
\end{quote}
Other outliers in the data also piqued participants' interest, such as those in their self-tracked health surveys or memories of notable air quality events:
\begin{quote}
    \textbf{P6}: I'm really looking for any outliers in the [survey] results... Let's look at that big outlier on, I guess, in January.
\end{quote}
\begin{quote}
    \textbf{P1}:  It might be easy to jump to a day like that, where we know $4^{th}$ of July is going to have fireworks.  
\end{quote}

Seeing these outliers in their data motivated participants to pause or modify their initial interview goals in favor of investigating the underlying events that caused them. They speculated about what might have led to the data feature:
\begin{quote}
    \textbf{P3:} That [spike] was a huge fire in Spanish Fork.  You could smell it everywhere, it was terrible.
\end{quote}
\begin{quote}
    \textbf{P5:} That's pretty interesting. [These spikes] make me wonder if it was all through the house, the fact that they're pretty similar. 
\end{quote}
Participants then asked to contextualize the data features with additional data streams to help validate their speculations or better understand what they were seeing. For example, both P2 and P3 wanted to use their text annotations to understand which behaviors were influencing the indoor air quality spikes they noticed:
\begin{quote}
    \textbf{P2:} I'd  be curious to see what our annotations are on the highest spikes... think you can do that?
\end{quote}
\begin{quote}
    \textbf{P3:} Being able to match my annotations with the larger spikes would be helpful to find patterns.
\end{quote}
For P5, however, reviewing his data uncovered unexpected late-night spikes that emerged during the last week of his deployment, and he requested a different type of contextualizing data stream:     
\begin{quote}
    \textbf{P5:} Can you overlay the temperatures? Outside, like the outside temp?
\end{quote}
P5's interest in incorporating his outside temperature measurements was a proxy for determining whether these periodic spikes may have been due to his furnace kicking on and off during cold temperatures.

As we analyzed participants' interviews, we began to recognize a recurring pattern regarding the sorts of data elements that caught their attention, and their method of investigating these elements. The participants would: 1) discover a prominent feature in their data, either through visual inspection or reflecting on their experiential knowledge; 2) determine if the feature warranted further inspection, and if so, attempt to correlate it with features in other data streams; and 3) speculate about underlying factors and potential behavior changes.  

In her data engagement interview, participant P3 distinguished between health- or air quality-motivated workflows, yet ultimately described the same exploration pattern:
\begin{quote}
    \textbf{P3:}  I mean this is an air quality thing, right? So if it were me, I would start with the air quality. When am I having a spike? I would then look at the spikes, and then from that I would try to correlate: On those days, what happened to cause a spike? Was anybody ill? Did we have an asthma attack? And try and do that. I think that's the direction I would go, because I'm thinking of it as an air quality thing. If it were sick thing [where] I'm trying to make [my son] healthier, I think I would start with his asthma data, and then go the other direction. So I think it just depends on which approach I would take.
\end{quote}

This exploration pattern naturally emerged in all of our participant interviews, regardless of their prior preparation or motivation, and independent of whichever particular feature they chose to enter their personal data. Air quality measurement spikes, key dates, personal associations or memories, text annotations, and survey response outliers were the most frequent subjects that drew the participants' attention in their interviews.

\subsection{Playful engagements} \label{sec:playful_engagement}

Our longitudinal study enrolled parents struggling to control the symptoms of severe asthma in their children and themselves. They participated because of their hopes that data about the air quality in their homes could improve the health of the asthmatics in their households --- a serious undertaking with clear implications for the health and well-being of their families. Nonetheless, their engagements with the deployed system, and with their data, led to numerous moments of play. 

For example, some of the participants' self-tracked annotations captured how repetitive data entry gave way to playful antagonism toward their deployments and data entry. In our interview with P4a, she admitted to her increasingly deprecating annotations about her mom:
\begin{quote}
    \textbf{P4a:} Towards the end I think I got a little more joke-y with it -- ``oh you know, mamma’s been cooking again.'' Before it was ``mom burnt the chicken.'' Now it's ``nothing new is happening now!''
\end{quote}
Similarly, P1 explained why some of her annotations personified the deployed system's alert mechanism:
\begin{quote}
    \textbf{P1:} We did start putting snarky remarks in some of the comments [laughs]. You probably noticed! There was a while where my oven had burnt pizza on the bottom, and every time we turned the oven on [the alert] was like "HEY!! HEY!! HEY!!!" And I was just like ``yep, still haven't cleaned my oven!  You want to come clean my oven?  'Cause I still haven't cleaned my oven.''  [laughs]
\end{quote}

Cooking annotations were prominent in each participants' dataset and occupied the majority of discussion around household behaviors, with a special focus on bacon:
\begin{quote}
    \textbf{P2:} That's me, burning the bacon.
\end{quote}
\begin{quote}
    \textbf{P5:} Oh, and you have where I put in that we were cooking bacon: ``Me cooking bacon.``
\end{quote}
\begin{quote}
    \textbf{P3:} We did the same thing over and over again. Bacon. [laughs] I think that 90\% of our annotations are probably bacon. I like bacon! [laughs]
\end{quote}
P3 was especially amused to see how often bacon was present in her data, bringing it up multiple times throughout her data engagement interview. These amusements turned into deeper engagements with her data:
\begin{quote} 
    \textbf{P3:} I'm looking at [my data] and there's a \textbf{\textit{lot}} of bacon ... If I notice that every single spike is because we're cooking bacon, then I might think, is that a problem? What is it that cooking bacon puts in the air? Is that a bad thing, or, the smell's a good thing, right? It's a good thing.
\end{quote}

Unlike most of the participants, P2 was much more difficult to engage in the data review process due her lack of confidence in her analytic abilities, and a stated preference for having medical professionals interpret her information.  As the most severe asthmatic in our study, P2 was also overly cautious to make any behavioral changes due to the perceived consequences of incorrectly interpreting her data. 
These facts made it all the more surprising when P2 \textit{also} got drawn into her data through exploring her spikes and cooking annotations.  Like others, she, too, began cracking wry jokes about burning food and cooking bacon:
\begin{quote}
     \textbf{P2:} That's me, burning the bacon ... I'm real good in the kitchen, I can tell you that!
\end{quote}

Every one of our data engagement interviews captured moments of play. Participants' interpretations of what they were seeing in their data, filtered through their self-deprecating, sarcastic humor, revealed insights into the challenges of balancing health with other priorities:
\begin{quote}
    \textbf{Interviewer:} what's interesting for you here?\\
    \textbf{Husband:} ...my painting\\
    \textbf{P1}: You were spraying [figurines] on the 23rd... Being a geek is hazardous to your health!\\
    \textbf{Husband}: Sorry, had a D\&D coming up.
\end{quote}

These exchanges illustrate the overall playfulness we witnessed when engaging people with their data.  This sense of play that emerged when exploring personal data highlights a broader dynamic within each participant's deployment. Our existing rapport, plus the potential for seeing their data in new ways, made participants excited to dive into their data and learn new things.  This excitement manifested itself differently for each participant, but a playful sense of curiosity helped pull people into their data and unpack what they had to show. 

\subsection{Serendipitous discoveries} \label{sec:serendipity}

Participants' concurrent enrollment in a  national asthma study \cite{PRISMS} and our own visualization design experience primed us to assume that they would be goal-oriented when it came to engaging their data at each stage of the longitudinal study. Much of what we observed in the data engagement interviews, however,  was productive free-form exploration, often facilitated by playful engagements. 

For example, P3 used her annotations as a way to self-experiment with her cooking habits to uncover which kind of cooking oil produced the fewest spikes during her deployment \cite{Moore2018}.  During her data engagement interview, however, her playful interest in bacon gave way to a broader exploration of her cooking habits. This exploration led P3 to a different view of activities that affected her indoor air quality:
\begin{quote}
    \textbf{P3:} I remember making the connection between the olive oil and the [spikes]. And I also knew that it was kind of every time we cooked bacon there was a [spike]. But I guess I didn't realize how many of them, overall, were actually cooking episodes... Like ``cooking pancakes'', ``cooking eggs'', ``[my daughter] burning the tortillas''. It's all cooking.
\end{quote}

Most other participants also experienced similar realizations after idly exploring their data. P4 came unprepared for her interview without considering what she wanted to explore ahead of time.  Yet, when reviewing her data during the interview, unexpected spikes caught her attention: 
\begin{quote}
    \textbf{P4:} It seems like most of [the spikes] are actually in [P4a's] bedroom, which surprised me.
\end{quote}
In contrast, P5 came to the interview with the goal of finding connections between his indoor air quality and respiratory health. However, unexpected spikes distracted him:
\begin{quote}
    \textbf{P5:} Oh, and I have no idea what would be in the room making it that high.  Why would there be a spike in the bedroom, and not downstairs?
\end{quote}
Following this discovery, he proceeded to spend over half of the interview attempting to find possible sources of his mysterious, nightly indoor air quality spikes. 
P2, who was the least willing to engage with her data, was also drawn in by reviewing spikes and their annotations. She saw that many of her air quality spikes were from cooking, and she became unexpectedly invested in understanding the extent of cooking spikes in her data:
\begin{quote}
    \textbf{P2:} I bet that [spike] is cooking, too.  If it's not I'm going to be surprised.
\end{quote}

The data engagement interviews gave participants the time and space to stumble upon unexpected and surprising observations, often leading to new insights. 
Despite our attempts to conduct goal-oriented analysis --- by explicitly priming participants ahead of each interview to discuss their analysis goals --- the most productive outcomes came from serendipitous discoveries. 

\subsection{Reluctance to personally analyze} \label{sec:reluctance}

Participants' levels of engagement during their interviews --- and throughout the longitudinal study --- were as varied as their questions. Some participants enthusiastically engaged with their data, some were reluctant but still tenacious, and others were difficult to motivate at all. Regardless of their level of engagement, however, every participant stopped short of advocating for a tool that would allow them to analyze their data in similar ways as the data engagement interview.  We asked each participant how likely they would be to use an idealized tool that offered the same flexible features.  Other than the teenage participant (P4a) who felt she ``might use it,'' all other participants were pessimistic: 
\begin{quote}
    \textbf{P1}: As a busy mom with small children, I just don't have the time.
\end{quote} 
\begin{quote}
    \textbf{P6}: I could do it, I just don't know how often I would.
\end{quote}  
\begin{quote}
    \textbf{P2}: I don't know that I would ever just pull it up and look at it for data's sake... I can take it back to my doctor.
\end{quote}

Participants' reluctance to engage with personal data is consistent with findings in other informatics disciplines. P2's preference for having her personal health data interpreted by medical professionals echos similar findings in other chronic health management research \cite{mamykina2008mahi}, and recent work shows even Division 1 collegiate athletes,  with access to vast stores of personal data and dedicated analysts, also resist engaging with their data for various reasons \cite{kolovson2020personal}. Both these contexts bear a resemblance to our own participants' circumstances as asthmatics living in a region that experiences some of the worst air quality days in the world \cite{slcAir}, lending evidence to broader, more complex challenges.

\section{Discussion}\label{sec:discussion}

We have outlined the \gap through a careful review of personal informatics and visual analytics literature.  This review shows how  personal informatics research leverages people’s deep contextual knowledge to collect and reflect on personal data, but requires further study to develop versatile analysis tools that empower them to engage with their personal data in unique ways; how visual analytics excels at designing flexible data analysis tools for trained domain experts working in professional contexts, complicating their transferability to personal contexts and less experienced users; and how the democratized analysis goals of everyday visualization seek to empower people in exploring data, but whose field has yet to design for flexible, in-depth analysis of personal data.
The complementary strengths of these three fields points the way toward opportunities for designing new tools and developing new methods that bridge the \gapNS. 

Reflecting on our experiences exploring the \gap reveals a number of visualization design opportunities: 
entry points as an immediate design recommendation for facilitating engagement with personal data; a play-based approach to challenge the traditional goal-oriented nature of visualization design in personal contexts; and a reconsideration of whether designing new tools is even the most productive way to support people in analyzing their personal data. We offer these ideas as a starting point for thinking differently about how we design for personal contexts, and to highlight the opportunities for the visualization community to make an impact in this space.
  
\subsection{Design for entry points}

In  \autoref{sec:patterend_engagement}, we discuss the common exploration patterns that we observed throughout the data engagement interviews.  Participants often fixated on the same kinds of visually prominent features within their plotted data  --- like spikes or outliers --- as well as performing the same exploration tactics when engaging with their data, in ways that bear a resemblance to established sense-making process models of intelligence analysts \cite{pirolli2005sensemaking}. Because participants shared these similarities regardless of their prior preparation or general interest in reviewing their data, designing for these tendencies could provide valuable insights into ways designers could make interfaces more engaging, especially in personal contexts.

These engagement and exploration patterns are examples of what the design community calls \textit{entry points} \cite{kirsh01work}, which are ``a point of physical or intentional entry into a design'' \cite{lidwell2010universal}. Rogers \etal identify the utility of entry points for interactive interfaces, recommending that designers incorporate them into their designs as a way to ``think about the coordination and sequencing of actions and the kinds of feedback to provide in relation to how objects are positioned and structured at an interface'' \cite{rogers2004new}. As an invitation to action, entry points bear a resemblance to affordances \cite{gibson1986ecological, norman1999affordance}. In the same way that affordances indicate potential interaction mechanisms to users, entry points provide hints for what can be done with engaging data \cite{thorvald2006triggers}.

As a concept, entry points describe an intuitive way for inviting users into a system or interface. Google Maps, for example, uses a device's location to generate an initial view as a way to experience its interface. 
Recent visualization tools explicitly implement entry points for investigating how people explore interactive visualizations \cite{Blascheck2019}, as well to facilitate rich conversations and exploration, encourage engagement with data, and make visualizations or datasets feel relevant to a wider audience \cite{walny2020pixelclipper}.
These approaches serve to help orient and engage people with unfamiliar tools or data sets \cite{hornecker07}, but the visualization community has largely overlooked entry points as a formalized construct to engage people with their personal data.

Existing theoretical work on entry points in the design community details how to design entry points into a system.
The design literature recommends establishing ``points of prospect'' to give an overview of the different ways people can engage with their data, and ``progressive lures'' to incrementally bring them into their data \cite{lidwell2010universal}.  This approach can lower barriers to entry and invite progressively deeper inspection.   Data engagement interviews \cite{Moore2021} can also benefit from considering entry points as a design element by helping to prioritize which analytic capabilities researchers should support. 

Human-computer interaction research into entry points can also help with determining the kinds of entry points different types of personal data may support. Tailoring interactions that reflect how people think of a particular data source may help lure them in to freely explore, and improve their overall chances for finding interesting parts of their data. For example, Choe \etal support basic temporal cuts for reviewing logged physical activity data within their Visualized Self web application \cite{choe2017understanding}, which users in their study found helpful and compelling. Time Lattice \cite{miranda2018time} generalizes this concept to support interactive analysis and comparisons for large-scale and distributed sensor data over a broader set of user-selectable constraints.
    
Entry points like these can allow for more meaningful exploration of self-tracked data in the context of supporting comparisons for how personal data vary across specific constraints or conditions, whether these are temporal (\textit{Q: What time of year is typically worst for air quality?}), geographic (\textit{Q: What part of the state has the best air quality?}), relative (\textit{Q:What's the air quality like when other people are sick}), absolute (\textit{Q: How many days were above the red air quality cutoff?}), or a mixture of these, and more. Systems supporting these approaches stand to better reflect how people think of their data and their lives, especially when compared to the static, linear time series plots typically provided for reviewing personal data.

The ability to seek or see one's self in data plays a significant role how people engage with and experience visualizations \cite{peck2019data}. More generally, any guidelines for developing engaging visualization systems will require a better understanding of what compels someone to start exploring visualizations of their data.  As a concept, entry points offer a promising approach to help address the shortcomings behind the \gapNS. Understanding what users find interesting or engaging in their data can help designers lower barriers and improve usability by  identifying compelling entry points into those datasets or visualizations.  Furthermore, researching entry points in personal contexts can help formalize ways people engage with personal data, along with common priorities, interests, or data characteristics they find especially relevant across different use-cases.
    
\subsection{Design for play}

We approached the data engagement interviews with a goal-oriented mindset, expecting our participants to come prepared with their own goals and approaches for analyzing their data. We asked participants to think about what they wanted to know about their data ahead of their interviews, and we structured the start of the interview to probe into their specific questions. Despite our attempts to prime our participants, some were disinterested in defining a goal and digging into their data as we detail in  \autoref{sec:reluctance}.

This lack of enthusiasm may have posed problems for more traditional, retrospective think-aloud interview methods, but the opportunity to directly engage with their personal data in the data engagement interview led each participant --- even those who \textit{were}  prepared with analytic goals --- to quickly distract themselves in playful ways as we guided them to examine their data. As we discussed in  \autoref{sec:serendipity}, participants preferred to freely explore their data, but not in the ways the visualization literature describes data exploration.  Although Brehmer \& Munzner's task typology \cite{brehmer2013multi} defines exploration as ``searching for characteristics'', the explorations of our participants had less to do with clear searching objectives, and more to do with stumbling into serendipitous discoveries. We had attempted to promote a goal-oriented experience, but this playful and open-ended exploration was what kept people engaged with their data and what motivated them to proceed through their interviews. 

This type of playful engagement is traditionally overlooked in visualization research in favor of more goal-oriented behavior.  The visualization community's focus on goals perhaps comes as a consequence of a decades-long framing of visualization as a vehicle for cognitive amplification and insight generation \cite{card1999readings}. Fun and enjoyment are secondary considerations, if they are considered at all. Prior work on ways to design for fun further codify this prioritization with recommendations to consider fun and enjoyment only after ``the functionality and usability have been accommodated in the design``\cite{shneiderman04fun}.  Our findings provide evidence that people engage with their data in productive ways through play, and lends support for prioritizing play as a first-class design requirement. 

Play, and the dynamics that give rise to or influence play, have been studied at length in human-computer interaction and design fields. Prior work shows people are likely to engage in activities seen as fun or playful, such as interactive filters on social media \cite{dodoo2021snapping}, or experiences that elicit curiosity from their ambiguous or unspecified outcomes \cite{tu2020curioscape, curiosityCube}. 
Designs that prioritize play and entertainment lead to more engaging experiences \cite{Deterding2011situated}, as do interfaces that are designed through processes that include situated play design \cite{bertran2019chasing}, ludic design \cite{gaver2004drift}, and playification \cite{scott2014playcentric}. 
Bertran \etal describe how \textit{situated play design} can inform early aspects of a design process by identifying design targets and inspiration based on their capacity for playful experiences \cite{bertran2019chasing}. This design approach also recommends engaging users within these contexts to explore how their play is integrated with their activities, and to use these dynamics as design targets that can be further refined with iterative prototyping \cite{dow2009efficacy}.  Elements of ludic design similarly recommend promoting curiosity and minimizing externally defined goals as a way to embrace ambiguity and break away from the requirements of meeting participants' every need \cite{gaver2004drift}.  Playification \cite{scott2014playcentric, thibault2017play} advocates designing for playful and intrinsically compelling experiences over more extrinsically gamified elements, like scores, that apply game mechanics to nonplayful activities \cite{deterding2011game}. 
    
Our observations that participants were naturally given to playful engagements and open-ended exploration when engaging their data, and reluctant to engage with an extrinsically goal-oriented analysis tool, resonates with the motivations for play-based design processes.

We speculate that characterizing people's naturally playful behaviors can help inform design elements for developing more engaging experiences in personal contexts.  This focus, however, requires that the visualization community re-evaluate its goal-oriented design bias and explore what playfulness means within personal contexts.  What is fun in the context of exploring personal data, and what does it mean to make something fun \textit{enough} to engage users?  This framing highlights alternative motivations that prioritize fun and enjoyment as first-class design criteria.
        
\subsection{Reconsider designing tools}

In our previous work, we provided evidence that data engagement interviews were productive and allowed our participants to learn new things from their data \cite{Moore2021}. In  \autoref{sec:playful_engagement} and \autoref{sec:serendipity} of this paper, we further validate this interview method with additional evidence on how people were able to playfully engage with their data and explore in open-ended ways to arrive at serendipitous discoveries, regardless of their level of preparation or interest. We also described in \autoref{sec:diverse_goals} that the interviews exposed a set of diverse goals and questions that our participants had, complicating the design space for potential visual analysis tools. Furthermore, in \autoref{sec:reluctance}, we detail how the majority opinion among our participants was that they were unlikely to analyze personal data on their own, from a lack of time (P1), interest (P6), relevance (P3, P4, P5), or confidence (P2).  Finally, although we succeeded at engaging participants with their data, we were unsuccessful at focusing participants' data engagements toward goal-oriented data analysis. Instead, efforts to encourage goal-oriented review fell apart, with participants engaging in open-ended exploration once they got into their data. This finding suggests that any tool designed for our participants would need to be very flexible, but not overly complicated; and even if we developed such a tool, it remains unclear whether our participants would be motivated to use it. 

Designing tools to support people engaging with personal data is a difficult and time-intensive process, and requires deep, contextual knowledge for reaching reliable or meaningful interpretations \cite{tolmie2016has}.  These requirements raise doubts about how a designer can expect to know what is important within people's data or the ways it can relate to their diverse goals.  These concerns are further compounded when any applied analysis may inadvertently obscure or remove information that users may find relevant or important \cite{orcutt1968data}. 
Furthermore, if designing for personal data is difficult from its reliance on personal knowledge, and the people who have this personal knowledge are the same ones who -- for various reasons -- cannot or will not afford the time to engage with their data, where does this leave us? If asthmatics living in an area that experiences some of the worst air pollution in the world \cite{slcAir} cannot be motivated to engage with their personal air quality data, what hope do we have of mobilizing \textit{anyone} in open-ended personal visual analytics? 

We question whether it makes sense to expend our time and resources to develop potentially complex and design-intensive tools for small user groups that often struggle to translate these tools to other contexts.  Taken to extremes, a truly generalized analysis tool capable of approaching these requirements runs the risk of reinventing Tableau or Excel --- an overwhelming design proposition in its own right. Given the diverse breadth of participants' questions throughout our longitudinal study, we pose the argument that any standalone visual solution may be impractical or impossible to provide.

In spite of these challenges, our data engagement interviews were especially productive at engaging people in exploring their data. This observation leads us to ask: What if the solution is not another tool, but something wholly different? What if the antidote to increasingly sophisticated and customized visual analysis tools is an investment in communal systems or infrastructures that allows people to cooperatively share their data with dedicated data experts? These social systems could help offset the intellectual burdens of traditional visual analytics systems and enable those experienced with sophisticated tools to collaborate alongside people with personal data to leverage the strengths of each community: computation and context. This idea is not new, and previous work has outlined some logistics and challenges of this approach \cite{heer08collaborative}, yet more work is needed to understand ways to attract and sustain professional attention; motivate people to volunteer their time and expertise; and manage resources, systems, and structures that could make this a reality. 
    
We cannot design new ways to support people engaging with their data using the same thinking that created the \gap in the first place. Instead, this gap offers an opportunity for the visualization community to develop new methods for exploring the gap and new designed futures to bridge it. 

\section{Conclusion }\label{sec:conclusion}

This work draws attention to the \gap through a review of personal informatics and visual analytics literature and our own experiences working with asthmatics over a multiyear indoor air quality sensing project.  The gap describes a lack of attention to flexible analytic systems for supporting people engaging with their personal data, and could benefit pulling from the research strengths of personal informatics, visual analytics, and everyday visualization fields: the strength of personal informatics research for developing systems that support users to collect and reflect on personal data; the strength of visual analytics research for developing flexible and customized analysis tools; and the strength of everyday visualization research for empowering people to explore data in personal contexts.
    
Designing within the \gap requires new thinking and approaches compared to existing methods.  We believe this work points to a previously untapped research space with a wide rage of design opportunities for the visualization community, and one that stands to be approached from multiple perspectives. Broadening design thinking beyond rigid, goal-oriented tools to consider alternative priorities like designing for exploration, play, and more collaborative systems can leverage the strengths and skills of analytic professionals to help people get the most from their data. Acknowledging the \gap exists --- and taking steps to explore it --- is an important first step in addressing the present limitations for gaining deeper insights into what people want to do with their data. We offer this work as call to the community to explore the \gap through careful, qualitative work that unpacks personal data engagement and explores design opportunities that challenge normative visualization design.

\vspace{-2mm}
\section*{Acknowledgements}
\vspace{-2mm}

Many thanks to: our study participants for sticking with us all these years; Greg Furlich for his indispensable real-time data analysis; the Visualization Design Lab members for constructive criticism on multiple drafts and research pitches; and our reviewers for their thoughtful and detailed feedback.   

This work was supported by the National Institute of Biomedical Imaging and Bioengineering of the National Institutes of Health under Award Number U54EB021973. The content is the sole responsibility of the authors and does not necessarily represent the official views of the National Institutes of Health.

\bibliographystyle{unsrt}
\bibliography{references}

\end{document}